\documentstyle[12pt]{article}
\begin{document}
\begin{flushright}
October 1995\\
\end{flushright}
\vspace{2em}
\begin{center}
\renewcommand{\thefootnote}{\fnsymbol{footnote}}
{\large \bf Newman - Penrose Formalism for Gravitational Shock Waves}\\
\vspace{4em}
Koichi HAYASHI${}^{1}$ and Toshiharu SAMURA${}^{2}$
          \footnote[2]{Electric mail address:
samura@jet.earth.s.kobe-u.ac.jp}\\
\vspace{3em}
${}^{1}${\it Department of Mathematics and Physics,}\\
        {\it Faculty of Science and Technology,}\\
        {\it Kinki University, Higashi-Osaka,}\\
                {\it  Osaka 577, Japan}\\
\vspace{2em}
${}^{2}${\it The Graduate School of Science and Technology}\\
                {\it Kobe University, Nada, Kobe 657, Japan}\\
\end{center}
\vspace{2em}

\begin{center}
\begin{abstract}
\baselineskip 12pt

First order perturbations for the fields with spin on
the background metric of the gravitational shock waves are discussed.
Applying the Newman -- Penrose formalism, exact solutions of
the perturbation equations are obtained.
For particle physics, this would be one approach to the problem of scattering
paticle at Planck energy.

\end {abstract}
\end{center}
\vspace{3mm}
\clearpage
\baselineskip 14pt

\section{Introduction}

The gravitational shock wave (GSW) of a black hole is a solution
that is  obtained by the black hole moving at the limit of light
velocity.
Recently, we have calculated the metrics both for  Schwarzschild and Kerr
black holes \cite{HS94}.
Comparing with the GSW metric derived by Aichelburg and Sexl
\cite{AS71}(below AS), our metrics are interesting for the point that the
mass of the black hole is finite.

As physical applications of these metrics, two cases have been considered:
the gravitational waves emission when two black holes collide \cite{DP92},
and the scattering of particles at Planck scale as a model of quantum
gravity \cite{QG}.
In these papers, these applications had been calculated using only the AS
metric.
We have also investigated these using our metric \cite{HS94b}.

In this paper, we apply the Newman -- Penrose (NP) formalism
for GSW metrics and calculate various physical quantities in perturbations.
The usual approach to obtain perturbed solution is
perturbing the metric directly and solving for the resulting perturbed
field equation. However, it is in general very difficult to obtain
the direct solution for metric perturbation equation
except for simple cases.

On the other hand, the approach of the NP formalism provides soluble
perturbation equations for much wider cases. The perturbations of the
complicated metrics like
the Kerr black hole or Reissner -- Nordstr\"om one are successful
by this approach. In the NP formalism, the perturbation equations for
Weyl tensors
decouple to independent equations, for which the partial wave analysis
are posible using suitable radial
functions and spin weighted spherical harmonics.

When perturbed solutions are obtained, we can deal with
astrophysical applications: e.g. stability of a black hole,
tidal friction effects, superradiant scatterings, and gravitational
wave
processes.
The NP formalism of the GSW metrics derived in this paper is not
similary
another mathematical method to obtain the results by ordinary
procedures,
but the motivations which would shed new lights on physics exist.
Especially, this approach is suitable to treat the
scattering problem at the Plankian energies.
Moreover, we can challenge to several unsolved
problems by this method: the scattering of particles with spin,
the collision of two GSW's,
the scattering of a particle moving near a  black hole, the
spontaneous emission of radiation, and so on.

In this paper, we show the first step to these approaches.
In section 2, we choose a simple NP tetrad of GSW metric,
then, the spin coefficients, the Wely tensors, the Ricci tensors
and the scalar
tensor are calculated.
The field equation of a scalar field in flat space--time is given by
the Klein -- Gordon equation, $(\Box + m^2) \phi=0$.
In the curved space--time, this is modified to
$( \nabla_\alpha \nabla^\alpha + m^2) \phi =0$,
using covariant derivatives.
For the metric of GSW, the solution of this equation is given by
't Hooft \cite{QG}, and we can see the effects of GSW metric on the
behavior of the scalar field.
How about the other fields with spin?
This is the main subject of this paper.
We calculate the effects of GSW metric on various fields with spin
by perturbations.
In section 3, we derive the differential equations
for a test nertrino field (spin $= 1/2$), a test electromagnetic
field
(spin $= 1$) and a gravitational perturbation (spin $= 2$) in the
background metric of GSW, Exact solutions of these equations
are given. Section 4 is devoted
to conclusious and discussions.

\section{Newman -- Penrose Formalism}

In the Newman-Penrose (NP) formalism, a set of null tetrad
{\boldmath$\ell$}, {\boldmath$n$}, {\boldmath$m$}
and {\boldmath${\overline{m}}$} is to be introdeced, where
{\boldmath$\ell$}, {\boldmath$n$} are real, and
{\boldmath$m$}, {\boldmath${\overline{m}}$} are complex
conjugates of each other.
These must satisfy the orthogonality conditions:
\begin{eqnarray}
\mbox{\boldmath$\ell$}\cdot\mbox{\boldmath$m$}~=~
\mbox{\boldmath$\ell$}\cdot
\mbox{\boldmath${\overline{m}}$}
{}~=~\mbox{\boldmath$n$}\cdot\mbox{\boldmath$m$}~=~
\mbox{\boldmath$n$}\cdot
\mbox{\boldmath${\overline{m}}$}
{}~=~0~~,
\label{eqn:1}
\end{eqnarray}
null conditions:
\begin{eqnarray}
\mbox{\boldmath$\ell$}\cdot\mbox{\boldmath$\ell$}~=~
\mbox{\boldmath$n$}\cdot\mbox{\boldmath$n$}
{}~=~\mbox{\boldmath$m$}\cdot\mbox{\boldmath$m$}~=~
\mbox{\boldmath${\overline{m}}$}\cdot
\mbox{\boldmath${\overline{m}}$}
{}~=~0~~,
\label{eqn:2}
\end{eqnarray}
and the normalization conditions:
\begin{eqnarray}
\mbox{\boldmath$\ell$}\cdot\mbox{\boldmath$n$}~=~1~~
{\rm{and}}~~
\mbox{\boldmath$m$}\cdot\mbox{\boldmath${\overline{m}}$}~=~-1~~.
\label{eqn:3}
\end{eqnarray}

Now, we will investigate the NP formalism of the gravitational
shock wave (GSW) metrics which are derived by us \cite{HS94}.
For the construction of a null -- tetrad frame for the NP
formalism, we must find the null tangent vectors for geodesics.
The GSW metric is generally written by:
\begin{eqnarray}
ds^2~=~du~dv-d\rho^2-\rho^2d\varphi^2-A(\rho)~\delta\left(u\right)~du^2~~,
\label{eqn:4}
\end{eqnarray}
where $u=t-z, v=t+z$, and the concrete forms of $A(\rho)$ are
given in the previous paper. The null geodesic of (\ref{eqn:4})
satisfy:
\begin{eqnarray}
\frac{1}{2} \dot{u}~&=&~{\rm{constant}}~\left( =\frac{1}{2} \right)
\nonumber\\
\frac{1}{2} \dot{v}+A(\rho)~\delta(u)~\dot{u}
        ~&=&~{\rm{constant}}~\left( =0 \right) \nonumber\\
\rho^2 {\dot{\varphi}}~&=&~{\rm{constant}}~\left( =0 \right)
\nonumber\\
\ddot{\rho}~=~\rho {\dot{\varphi}}^2&-&\frac{1}{2}A'
\left( \rho \right) \delta \left( u \right) \dot{u}^2~~,
\label{eqn:5}
\end{eqnarray}
where dot and dash denotes the derivative with respect to the
affine parameter $\lambda$,
and $\rho$, respectively.
With no loss of generality, we take the constants of the RHS
of (\ref{eqn:5}) to the numbers in the parenthesis.
With this choice, $\dot{u}=1$, so that $u$ can be identified as
the affine parameter $\lambda$ itself.
The equations in (\ref{eqn:5}) are rewritten as
\begin{eqnarray}
\dot{u}~&=&~1\nonumber\\
\dot{v}~&=&~-2A\delta\nonumber\\
{\dot{\varphi}}~&=&~0\nonumber\\
\dot{\rho}~&=&~-\frac{1}{2}A'\theta~~,
\label{eqn:5-1}
\end{eqnarray}
where $\theta$ is the step function.
For simplicity we put $A(\rho)$, $\delta(u)$,
$\theta\left(u\right)$, and $\left.A' \left( \rho \right)
\right|_{\rho=\rho_0}$ as
$A$, $\delta$, $\theta$ and $A'$, respectively.

Then since the null tangent vector is
\begin{eqnarray}
v^i~=~\left(\dot{u},~\dot{v},~\dot{\rho},~\dot{\varphi}\right)
{}~=~\left(1,~-2A\delta,~-\frac{1}{2}A'\theta,~0 \right)~~,
\label{eqn:6}
\end{eqnarray}
the vector {\boldmath$\ell$} of the NP formalism is taken as
$v^i$ itself:
\begin{eqnarray}
\ell^i~&=&~\left(\ell^u,~\ell^v,~\ell^{\rho},~\ell^{\varphi} \right)
{}~=~\left(1,~-2A\delta,~-\frac{1}{2}A'\theta,~0 \right)~~,\nonumber\\
\ell_i~&=&~\left(0,~\frac{1}{2},~\frac{1}{2}A'\theta,~0 \right)~~.
\label{eqn:7}
\end{eqnarray}
It is to be noted that $\ell^i \ell_i~=~-A\delta-(\dot\rho)^2~=~0$,
since the derivative of the equation ${\dot{\rho}}^2=-A\delta$
with respect to $\lambda$ gives exactly the last equations in
(\ref{eqn:5}).

For the null vector {\boldmath$n$} and {\boldmath$m$} we will
take
\begin{eqnarray}
n^i~&=&~\left(a~,b~,c~,d~\right)~~,\nonumber\\
m^i~&=&~\left(0~,\alpha~,\beta~,{\rm{i}}\gamma~\right)~~.
\label{eqn:8}
\end{eqnarray}
 From conditions of (1)(2)(3), $m^i$ is determined uniquely:
\begin{eqnarray}
\alpha~=~-\frac{1}{\sqrt{2}}A'\theta,~~\beta~=~\frac{1}{\sqrt{2}},
{}~~\gamma~=~\frac{1}{\sqrt{2}\rho}~~,
\label{eqn:9}
\end{eqnarray}
while for $n^i$, the following equations are obtained:
\begin{eqnarray}
&&b+A'\theta c~=~2 \nonumber\\
&&A'\theta a+2 c~=~0 \nonumber\\
&&c^2~=~ab+A\delta a^2~~.
\label{eqn:10}
\end{eqnarray}
The simplest solution is
\begin{eqnarray}
a~=~0,~~b~=~2,~~c~=~0~~.
\label{eqn:11}
\end{eqnarray}

Let us summarize the null vectors of NP formalism which
we have chosen:
\begin{eqnarray}
\ell^i~&=&~\left(1,~-2A\delta,~-\frac{1}{2}A'\theta,~0 \right)~~,
\nonumber\\
n^i~&=&~\left( 0,~2~,0~,~0 \right)~~,\nonumber\\
m^i~&=&~\left(0,~-\frac{1}{\sqrt{2}}A'\theta,\frac{1}{\sqrt{2}},
{}~\frac{\rm{i}}{\sqrt{2}\rho} \right)~~.
\label{eqn:12}
\end{eqnarray}
and corresponding covariant vectors are
\begin{eqnarray}
\ell_i~&=&~\left(0,\frac{1}{2},~\frac{1}{2}A'\theta,~0 \right)~~,
\nonumber\\
n_i~&=&~\left( 1,~0~,0~,0~ \right)~~,\nonumber\\
m_i~&=&~\left(-\frac{1}{\sqrt{2}}A'\theta,~0,~-\frac{1}{\sqrt{2}},
{}~-\frac{\rm{i\rho}}{\sqrt{2}} \right)~~.
\label{eqn:13}
\end{eqnarray}

The nonvanising spin coefficients (for example, see (286) in
Chapter 1 of
Chandrasekhar \cite{Chandra83}) are
\begin{eqnarray}
\kappa~=~\frac{1}{2\sqrt{2}}A'\delta,~~\rho~=~-\sigma~=~
\frac{1}{4\rho}A'\theta,
{}~~\alpha~=~-\beta~=~-\frac{1}{2\sqrt{2}\rho}~~.
\label{eqn:14}
\end{eqnarray}
Here, note that $A'$ is a constant.

Finally, we can calculate the components of the Wely tensor,
$\Psi_i$, Ricci
tensor, $\Phi_{ij}$, and scalar tensor, $\Lambda$
 (see (294) and (300) in Chap.1 of \cite{Chandra83}).
Then we see all components of these tensors vanish:
\begin{eqnarray}
\Psi_i~=~\Phi_{ij}~=~\Lambda~=0~~\left( i,j=0~\sim~4 \right)
{}~~\left( {\rm{For}}~u\neq0 \right).
\label{eqn:15}
\end{eqnarray}
Since the space -- time of GSW is empty except for $u=0$,
this is rather a natural result.
The whole space -- time is the patchwork of two flat
Minkowskii spaces pasted at $u=0$, where the continuity is
destroyed.
Because of this discontinuity, (\ref{eqn:14}) is obtained and
physics are not trivial even with (\ref{eqn:15}).
In the next section we consider the perturbation over this
background space -- time.
Although this is almost flat, it reflects this discontinuity
through (\ref{eqn:14}).

\section{Perturbations on the Gravitational Shock Waves
Background}

The NP equations are system of first -- order differential
equations linking the tetrad, the spin coefficients, Wely tensors,
Ricci tensors and the scalar curvature. The perturbed
geometries in NP
formalism are specified by:
\begin{eqnarray}
\mbox{\boldmath$\ell$}~=~\mbox{\boldmath$\ell$}^u+
\mbox{\boldmath$\ell$}^p,~~
\mbox{\boldmath$n$}~=~\mbox{\boldmath$n$}^u+
\mbox{\boldmath$n$}^p,~~
\mbox{\boldmath$m$}~=~\mbox{\boldmath$m$}^u+
\mbox{\boldmath$m$}^p,
\label{eqn:16}
\end{eqnarray}
where the supersuffix $u$ and $p$ means "unperturbed"
and "perturbed", respectively. All the NP quantities can be
written in
this form. The spin coefficients $\kappa$, $\rho$, $\sigma$,
$\alpha$
and $\beta$ have both the unperturbed quantities and the perturbed
ones. The other spin coefficients, all Wely tensors, Ricci tensors
and the scalar curvature have only the unperturbed quantities.
The
complete set of perturbation equations are obtained from the
NP equations
by keeping perturbed terms only to first order.

In this section, we will derive the
source free perturbation equations for two component neutrino
fields,
electromagnetic fields, and gravitational fields on the GSW
background.

\subsection{Neutrino equations}

In the NP formalism for the GSW space -- time, the Dirac
equations of the
massless particle are written by the following equations
(see (108) in
Chap.10 of \cite{Chandra83}):
\begin{eqnarray}
\left( D-\rho \right) F_1+\left( \delta ^\ast-\alpha \right)
F_2~=~0
\label{eqn:17}\\
\left( \delta-\alpha \right) F_1+\Delta F_2~=~0~~,
\label{eqn:18}
\end{eqnarray}
where $F_1$ and $F_2$ are 2--spinors.
When we consider the realistic problem of the scattering of a
massless neutrino off
the GSW background, the $F$'s can be treated as the first order
test fields.
$D$, $\delta$ and $\Delta$ are directional derivatives along the
basis
null vectors defined by:
\begin{eqnarray}
DF~=~F_{;\mu}~\ell^\mu~~,~\Delta F ~=~F_{;\mu}~n^\mu~~, \delta
F~=~F_{;\mu} n^\mu~~.
\label{eqn:19}
\end{eqnarray}
Eliminating $F_2$ from (18) and (\ref{eqn:18}), we obtain the
equation for
$F_1$:
\begin{eqnarray}
\left[ \Delta \left( D - \rho \right) - \left( \delta^\ast - \alpha
\right) \left( \delta - \alpha \right) \right] F_1~=~0~~,
\label{eqn:20}
\end{eqnarray}
where we have used the relation, $\Delta \left( \delta^\ast -
\alpha \right) -
\left( \delta^\ast - \alpha \right) \Delta =0$.

We want to find the solution for $F_i$ which have the form:
\begin{eqnarray}
F_i~=~{\rm{e}}^{ {\rm{i}} \left( k_u u +k_v v+ m \varphi\right)}
f_i \left( \rho
\right)~~.
\label{eqn:22}
\end{eqnarray}
where $k_u$ and $k_v$ are constants and m is an integer,
General solutions are
to be obtained by superpositions of them.
When $u>0$, the equation of $F_1$ is given by:
\begin{eqnarray}
f_{1,\rho,\rho}+\frac{1}{\rho} f_{1,\rho} +
\left[ \left( 4 k_u k_v -
k_v^2 A^{'2} \right) - \frac{\left(m - 1/2 \right)^2}{\rho^2}
\right] f_1
{}~=~0~~,
\label{eqn:23}
\end{eqnarray}
and the solution of it is:
\begin{eqnarray}
f_1 \left( \rho \right)~=~J_{m-1/2} \left( y \right)~~~
\left( \rm{for}~
u>0 \right)~~,
\label{eqn:24}
\end{eqnarray}
where $J$ is the Bessel function, $y=\Omega \rho$ and $\Omega =
\sqrt{ 4
k_u k_v - k_v^2 A'}$.

For $u<0$, $f_1(\rho)$ is similarly obtained:
\begin{eqnarray}
f_1\left( \rho \right)~=~J_{m-1/2} \left( y' \right)~~
\left( {\rm{for}}~u<0 \right)~~,
\label{eqn:25}
\end{eqnarray}
where $y'=\Omega'\rho$ and $\Omega'=\sqrt{4k_uk_v}$.

Next, we will give the $F_2$. From (\ref{eqn:18}) and (\ref{eqn:22}),
$f_2$, which is the $\rho$--direction field of $F_2$, is
represented by
$f_1$:
\begin{eqnarray}
2 \sqrt{2}{\rm{i}}k_v f_2 ~=~-f_{1,\rho}+{\rm{i}}A'k_vf_1 +
\frac{m-1/2}{\rho}f_1~~\left( {\rm{for}}~u>0 \right)~~.
\label{eqn:26}
\end{eqnarray}
Using the formula, $J'_\nu(z) =-J_{\nu+1}(z)+
\nu/zJ_\nu(z)$, $f_2$ for $u>0$ is given by:
\begin{eqnarray}
f_2 \left( \rho \right) ~=~ \frac{\Omega}{2\sqrt{2}{\rm{i}}k_v}
\left\{J_{m+1/2}
\left( y \right) +\frac{{\rm{i}}k_vA'}{\Omega}J_{m-1/2}
\left( y \right)
\right\}~~
\left( {\rm{for}}~u>0 \right)~~,
\label{eqn:27}
\end{eqnarray}
while for $u<0$,
\begin{eqnarray}
f_2 \left( \rho \right) ~=~ \frac{\Omega'}{2\sqrt{2}{\rm{i}}k_v}
J_{m+1/2}
\left( y' \right) ~~
\left( {\rm{for}}~u<0 \right)~~.
\label{eqn:27-2}
\end{eqnarray}

It is to be noted that these solutions are the exact solutions for
the neutrino fields.

\subsection{Electromagnetic and Gravitational Equations}

Maxwell equations in the NP formalism
of the GSW geometry can be written by (see (330)--(333) in Chap.1 of
\cite{Chandra83}):
\begin{eqnarray}
\left( D-\rho \right) \phi_2-\delta^* \phi_1 ~&=&~0
\label{eqn:28}\\
\left( \delta-2\alpha \right) \phi_2-\Delta\phi_1  ~&=&~0
\label{eqn:29}\\
\delta \phi_1 -\Delta \phi_0 +\sigma \phi_2 ~&=&~0
\label{eqn:30}\\
\left( \delta-2\rho \right) \phi_1-\left( \delta^* -2\alpha
\right) \phi_0  ~&=&~0~~,
\label{eqn:31}
\end{eqnarray}
where $\phi$'s are Maxwell field strengths.
With $\Delta
\times$(29)$-\delta^*\times$(30), the equation for
$\phi_2$ is given by the following equation:
\begin{eqnarray}
\left[ \Delta \left( \Delta - \rho \right) - \delta^\ast
\left( \delta - 2\alpha \right) \right] \phi_2~=~0~~,
\label{eqn:32}
\end{eqnarray}
As in the previous subsection, we put
\begin{eqnarray}
\phi_i~=~{\it{e}}^{ {\rm{i}} \left( k_u u +k_v v+ m \varphi\right)}
R_i \left( \rho
\right)~~.
\label{eqn:33}
\end{eqnarray}
Then (\ref{eqn:32}) is written by the following differential
equation for
$u>0$:
\begin{eqnarray}
R_{2,\rho,\rho}+\frac{1}{\rho} R_{2,\rho} + \left[ \left( 4 k_u k_v -
k_v^2 A^{'2} \right) - \frac{\left(m - 1 \right)^2}{\rho^2} \right] R_2
{}~=~0~~.
\label{eqn:34}
\end{eqnarray}
The equation is easily solved:
\begin{eqnarray}
R_2 \left( \rho \right)~=~J_{m-1} \left( y \right)~~~ \left( \rm{for}~
u>0 \right)~~,
\label{eqn:35}
\end{eqnarray}
as before. On the other hand, the solution for $u<0$ is
\begin{eqnarray}
R_2 \left( \rho \right)~=~J_{m-1} \left( y' \right)~~~ \left( \rm{for}~
u<0 \right)~~,
\label{eqn:36}
\end{eqnarray}
where $y$ and $y'$ are given by (\ref{eqn:24}) and (\ref{eqn:25}).
As $\phi_1$ and $\phi_2$ are related by (30),
$R_1$ is given by
\begin{eqnarray}
R_1 \left( \rho \right) ~=~ \left( - \frac{\Omega}
{2\sqrt{2}{\rm{i}}k_v} \right)
\left\{ J_{m}
\left( y \right) +\frac{{\rm{i}}k_vA'}{\Omega}J_{m-1}
\left( y \right)
\right\}~~
\left( {\rm{for}}~u>0 \right)~~,
\label{eqn:37}
\end{eqnarray}
while for $u<0$, it is:
\begin{eqnarray}
R_1 \left( \rho \right) ~=~ \left( -\frac{\Omega'}
{2\sqrt{2}{\rm{i}}k_v} \right) J_{m}
\left( y' \right) ~~
\left( {\rm{for}}~u<0 \right)~~.
\label{eqn:38}
\end{eqnarray}
Finally, $R_0$ is given from (\ref{eqn:30})
\begin{eqnarray}
R_0 \left( \rho \right) ~&=&~\left( - \frac{\Omega}
{2\sqrt{2}{\rm{i}}k_v}
\right)^2 \left\{ J_{m+1}
\left( y \right)+\frac{2{\rm{i}}k_vA'}{\Omega}J_{m}
\left( y \right)
\right. \nonumber\\
&&~~~~~~~~~~~~~~~\left. +\left( \frac{{\rm{i}}k_vA'}
{\Omega} \right)^2 J_{m-1}\left( y \right)
\right\}~~\left( {\rm{for}}~u>0 \right)~~\\
\label{eqn:39}
R_0 \left( \rho \right) ~&=&~\left( - \frac{\Omega'}
{2\sqrt{2}{\rm{i}}k_v}
\right)^2 J_{m+1}
\left( y' \right) ~~
\left( {\rm{for}}~u<0 \right)~~.
\label{eqn:40}
\end{eqnarray}

Now we want to discuss about the first -- order perturbations of the
gravitational field itself.
We find that four of the eight NP Bianchi identities provide the
following
linear homogeneous equations to first order in the perturbations
(see (321) in Chap.1 of \cite{Chandra83}):
\begin{eqnarray}
\left(\delta^* + 2\alpha \right) \Psi_3 - \left( D -\rho \right)
\Psi_4~=~0 \\
\label{eqn:41}
-\Delta \Psi_0 +\left( \delta -2\beta \right) \Psi_1 +3\sigma
\Psi_2~=~0 \\
\label{eqn:42}
-\Delta \Psi_1 + \delta \Psi_2 +2\sigma \Psi_3~=~0\\
\label{eqn:43}
-\Delta \Psi_2 +\left( \delta+2\beta \right) \Psi_3 +\sigma
\Psi_4 ~=~0\\
\label{eqn:44}
-\Delta \Psi_3 +\left( \delta +4\beta \right) \Psi_4~=~0
\label{eqn:45}
\end{eqnarray}
where $\Psi$'s are the perturbed Wely tensors. We can calculate
$\Psi$'s by the same procedure as neutrino and electromagnetic cases,
and the final forms of $\Psi$'s are:
\begin{eqnarray}
\Psi _n &=& \exp(ik_u u+ik_v v+im\varphi)~B^{4-n}~J_{m-n+2}
        (y')~~\nonumber\\
&&~~~~~~~~~~~~~~~~~~~~~~~~~~~~~~~~~~~~~~~~~({\rm{for}}~u<0)~~,\\
\label{eqn:46}
\Psi _n &=& \exp(ik_u u+ik_v v+im\varphi)~
        B^{4-n}\sum^{5-n}_{l=1} {}_4{\rm{C}}_l D^{l-1} J_{m-n+3-l}
(y)~~\nonumber\\
&&~~~~~~~~~~~~~~~~~~~~~~~~~~~~~~~~~~~~~~~~~({\rm{for}}~u>0)~~,
\label{eqn:47}
\end{eqnarray}
where
\begin{eqnarray}
B = -\frac{\Omega'}{2\sqrt{2} i k_v}~~, D = \frac{ik_v A'}{\Omega}~~,
\nonumber\\
\end{eqnarray}
and ${}_4{\rm{C}}_l$ represents the binomial coefficient.

\section{Conclusions}

We have derived the first order perturbations for fields with
spin on the background metric of GSW.
Exact solutions of fields are obtained.
 From these, we can find the behavior of the field crossing
the discontinuity of GSW at $u=0$.
Especially, we can find, for example, the refraction angles and
scattering cross sections of various fields near the black hole which
is moving with a relativistic speed.
For particle physics, this would be one approach to the physics
at Planck
energy.
These physics are now under investigations.


\begin{thebibliography}{0}

\bibitem{HS94}
K. Hayashi and T. Samura,
Phys. Rev. {\bf{D50}}, 3666(1994).
\bibitem{AS71}
P. C. Aichelburg and R. U. Sexl,
Gen. Relativ. Gravit. {\bf{2}}, 303(1971).
\bibitem{DP92}
P. D. D'Eath, Phys. Rev. {\bf{D18}}, 990(1778);
P. D. D'Eath, and P. N. Payne, ibid, {\bf{46}}, 658(1992);
{\bf{46}}, 675(1992); {\bf{46}}, 694(1992).
\bibitem{QG}
For example see,
G.'t Hooft,
Phys. Lett. {\bf{B198}}, 61(1987);
ibid, Nucl. Phys. {\bf{B355}}, 138 (1990);
D. Amati et.al.,
Phys. Lett. {\bf{B197}}, 81(1987);
C.O.Loust\'{o} and N.S\'{a}nchez,
Phys. Lett. {\bf{B232}}, 462(1989);
ibid, Int. J. Mod. Phys. {\bf{A5}}, 915 (1990);
ibid, Nucl. Phys. {\bf{B355}}, 231(1991);
ibid, Nucl. Phys. {\bf{B383}}, 377 (1992);
H. Verlinde and E. Verlinde,
Nucl. Phys. {\bf{B371}}, 246 (1992);
M. Fabbrichesi et. al.,
Nucl. Phys. {\bf{B419}}, 147 (1994);
V. Ferrari and M. Martellini, Nucl. Phys.
{\bf{B385}}, 604 (1992).
\bibitem{HS94b}
K. Hayashi and T. Samura,
Planckian Scatterings of Massive Particles
and Gravitational Shock Waves,
preprint KOBE-FHD-94-05 and hep-th/9405013 (1994).
\bibitem{Chandra83}
S Chandrasekhar,
{\it{The Mathematical Theory of Black Holes}}
(Clarendon Press, Oxford, 1983).

\end{thebibliography}
\end{document}